\documentstyle[preprint,eqsecnum,aps,epsfig]{revtex}
\begin{document}

\def\pa{\partial}
\def\ov{\over}
\def\non{\nonumber }
\def\beq{\begin{equation} }
\def\eeq{\end{equation} }
\def\beqa{\begin{eqnarray}}
\def\eeqa{\end{eqnarray}}

\title{Landau Levels in the noncommutative $AdS_2$}
\author{Roberto Iengo{}$^a$ and R. Ramachandran{}$^{b,c}$}

\address{${}^a$~International School for Advanced Studies (SISSA),
I-34013 Trieste, Italy\\
INFN, Sezione di Trieste\\
{\tt iengo@sissa.it} }

\address{${}^b$~The Abdus Salam ICTP, Trieste,Italy}
\address{${}^c$~The Institute of Mathematical Sciences, Chennai 600 113,
India\footnote{permanent address}   \\
{\tt rr@imsc.ernet.in}}

\maketitle
\begin{abstract}
We formulate the Landau problem in the context of the noncommutative
analog of a
surface of constant negative curvature, that is $AdS_2$ surface,
and obtain the spectrum and contrast the same with the Landau
levels one finds in the case of the commutative $AdS_2$ space.
\end{abstract}

\section{Introduction}

Noncommutative spaces have been of current interest with various motivations,
in particular they arise in the framework of M-theory and in interesting
settings  of string and branes \cite{banks}, \cite{connes}, \cite{seib}
(the bibliography is so vast that we do not attempt comprehensive
referencing).

A sector of the study of the physics in noncommutative spaces concerns
the exploration of the consequences for the quantum mechanics
of one particle \cite{jack}, 
\cite{N1}, \cite{NP}.

The physics in the noncommutative spaces is closely related to the
problem of a charged particle moving on a surface with constant magnetic field
giving rise to Landau Levels (see for instance \cite{dunne}). 
Hence it is interesting to study
Landau Levels by comparing the settings of commutative and noncommutative
spaces. This research has been carried out for the case of the plane
\cite{NP},\cite{gamboa},\cite{bellucci}, the sphere\cite{NP} and  the
torus \cite{poly}.

In this paper we consider the Landau Levels problem in the case of
a surface of negative constant curvature, that is $AdS_2$.
The commutative case has been studied in various papers, \cite{comtet},
\cite{antoine}, and also been extended to cover the case of the higher genus
Riemann surfaces, which can be realized by a tessellation of $AdS_2$
\cite{avron}, \cite{iengo}. We may note in passing that,
as higher genus Riemann surfaces appear as building blocks
of higher orders in string perturbation theory, this provides a further link
between $AdS$ spaces and string theory, besides the celebrated
relation with conformal field theories.

We first of all recall in Section II
the results on the commutative $AdS$ surface,
by making an explicit derivation, using appropriate complex coordinates
and giving the resulting eigenfunctions, eigenvalues and their
(infinite) multiplicity.
We may also recall that for higher genus Riemann surfaces one gets
the same spectrum but with a finite multiplicity dictated by the
Riemann-Roch theorem \cite{iengo}.

Then we give the algebraic formulation of the same problem, by
expressing the Hamiltonian in terms of the generators of $SO(2,1)$
and representing $AdS_2$ as an embedding of a surface in the
flat $(2+1)$-dimensional Minkowski space.

In the case of the Riemann surfaces a quantization condition 
for the magnetic field naturally emerges. Indeed, the wave function 
can be regarded as a differential form on the surface (which is in fact
the proper object of the Riemann-Roch theorem) and the requirement
of definite monodromy for transport along noncontractible loops
implies a quantization condition.
One can also require some periodicity properties of the wave function
 in the noncompact case of $AdS_2$, like it is common to use
peridodic boundary conditions in quantum mechanical problems 
on noncompact spaces, for instance on the plane. The results 
are actually the same, except one has to assume a quantized value
for the magnetic field.   

The algebric formulation of Section II
will allow us to properly define the analogous problem
in the noncommutative setting, Section IIIA.
The commutation relations
among the Minkowski space coordinates are taken to be the ones
of the $SO(2,1)$, and the appropriate Casimir is fixed in order
to define the embedding in this case, similarly to
the construction for the noncommutative sphere done
in ref\cite{NP}. The resulting setting is described by two
commuting $SO(2,1)$ algebrae. We have not attempted the construction
of noncommutative higher genus Riemann surfaces.

We have first of all to define the Hamiltonian for the noncommutative case:
we assume it to be formally identical to the one defined on the commutative surface.

The next issue concerns how to define the constant magnetic field.
Here we have studied two options.

In the first one, we fix the two
Casimirs of the two commuting $SO(2,1)$ algebrae, similarly to what
was done in ref\cite{NP} for the sphere. 
With this option the Hamiltonian for the noncommutative case may not
be formally the same as for the commutative surface, and the commutative
limit may require some care and adjustment of parameters appearing in the Hamiltonian, 
see ref\cite{NP}. Here, we see that this option can be in conflict 
with the requirement that a universal (commuting or noncommuting) form of the 
Hamiltonian makes physical sense.   

In the second option, we keep fixed one observable among a complete set
of mutually commuting ones. This observable is formally identical
to the magnetic field defined in the commutative case. In this case,  
retaining the same Hamiltonian, formally identical to the one
defined on the commutative surface, makes always physical sense, and
the commutative limit is straightforward. 

By using the representation theory we obtain the spectrum in both options,
Section IIIB and Section IIIC respectively.
This is done in general for all possible representations of the algebra to 
yield the spectrum. Requiring, in addition, quantization of the eigenvalues in order to
explore the possible noncommutative generalization of the features holding
for Riemann surfaces, implies retaining only a quite small subset of the levels.
Actually, the construction of the Landau Levels on the noncommutative version of the
Riemann surfaces remains so far a completely unsolved problem, despite announcements
which may have appeared in the title of some paper.

\section{Landau Levels in $AdS_2$ }

We consider a constant magnetic field on $AdS_2$, that is a magnetic field
proportional to the curvature. We can describe $AdS_2$ by using complex
coordinates $z,\bar z$ in the upper half plane
$y>0$ and taking the Poincare' metric $g_{z\bar z}=1/y^2$:
\beq
ds^2= { dx^2+dy^2 \over y^2}.
\label{metric}
\eeq

The relevant covariant derivatives are
\beq
\nabla  = \pa +{B \over z-\bar z} \hskip.5truecm
\bar\nabla = \bar \pa +{B \over z-\bar z} .\label{nabla}
\eeq
and
\beq
[\nabla,\bar \nabla] = B/(2y^2).
\eeq

We take the Hamiltonian as
\beqa
H &=& -2g^{z\bar z}(\nabla\bar \nabla+\bar \nabla\nabla)
  - B^2 \non \\
&=&-4g^{z\bar z}\nabla \bar \nabla+B(1-B) \non \\
&=&-4g^{z\bar z}\bar\nabla\nabla-B(1+B).
\label{hamiltonian}
\eeqa

Notice that, by taking into account the appropriate measure,
the operators $-g^{z\bar z}\nabla\bar \nabla$ and
 $-g^{z\bar z}\bar \nabla\nabla$ are both semipositive
definite, and therefore

\beq
H\geq |B|(1-|B|). \label{bound}
\eeq
We take  $B>0$,
since the case $B<0$ is obtained by interchanging $z$ and $\bar z$.

Consider the eigenvalue problem
$$
H\Psi = E\Psi.
$$
The lowest eigenstate, i.e. $E=B(1-B)$, is obtained as a solution
of $\bar \nabla \Psi_0 =0$.

By defining $\tilde \Psi_0 = g_{z\bar z}^{B/2}\Psi_0$ we see that
this means $\bar \pa \tilde \Psi_0 =0$. This state is not unique.
The different states can be labeled by the eigenvalues of
an  operator (that commutes with $H$), which we will, in the following, 
identify with a generator of $SO(2,1)$:
\beq
J_3=-{i\over 2}((1+z^2)\pa +(1+\bar z^2)\bar\pa +(z-\bar z)B).
\label{J3}
\eeq
The explicit form of the set of lowest level eigenstates is
\beq
\Psi_0^{(n)}= {(z-\bar z)^B \ov (i+z)^{2B}}\left( {-i+z \ov i+z}\right)^n,
\eeq
corresponding to the eigenvalues $J_3 = B+n$, with $n$ any nonnegative
integer.

We observe that under a holomorphic coordinate transformation
$z\to z'={az+b\over cz+d}$ the wave function $\Psi$ transforms like
a differential form of the kind $T^{\bar B/2}_{B/2}$.
That is, if $z'$ is another local coordinate on the surface
and the domain of $z'$ intersect the domain of $z$, we
first observe that $g_{z\bar z}dzd\bar z=g_{z'\bar z'}dz'd\bar z'$
and that 
$$
H'=UHU^{-1},
$$  
with $U=({dz\over dz'}\cdot{d\bar z'\over d\bar z})^{B/2}$.
Therefore the wave function in the new coordinates is related to the
wave function in the old ones by
$$
\Psi '=U\Psi,
$$ 
that is $\Psi ({dz\over d\bar z})^{B/2}$ is invariant.
It follows that $\tilde \Psi= (g_{z\bar z})^{B/2}\Psi$ 
transforms like a $T_B$ form.

Also, the wave functions for the excited levels can be expressed as 
(we abbreviate, $g\equiv g_{z\bar z}$) 
\beq
\Psi_l^{(n)} = g^{B/2-1}\pa g^{-(B/2-1)}\cdot g^{B/2-2} \pa g^{-(B/2-2)}
            \cdots g^{B/2-l}\pa g^{-(B/2-l)} \Phi,
\eeq
where $\Phi$ is a $T^{\bar B/2}_{B/2-l}$ form.
The eigenfunction equation requires
$\tilde \Phi= (g_{z\bar z})^{B/2}\Phi$, which is a
$T_{B-l}$ form, to be holomorphic.

In conclusion the wave functions of the operator 
$J\circ J=j(1-j)$ are in correspondence with $T_{j}$ 
holomorphic forms.

On a Riemann surface the Riemann-Roch theorem tells us that the 
dimensionality of the holomorphic $T_j$ forms is $(2j-1)(h-1)$
where $h$ is the genus of the surface and $j=B, B-1,\dots$.
Therefore, on the
Riemann surface this quantity must be an integer.
We expect this quantization to generalize to the possible 
noncommutative version of the Riemann surface. 

Further, requiring periodicity for transport along noncontractible
loops fixes $j$ to be integer (it can also be half-integer
if we only require periodicity up to $\pm$ signs, giving
rise to the ``spin structures'' of the half-forms,
which are familiar from perturbative String theory). 
Periodic boundary conditions are very natural for eigenvalue
problems and we can choose to require it even for the $AdS_2$
surface (which can be considered as a limiting case of a 
very large Riemann surface). 
Therefore we are led to consider the case of $j$ to be integer 
(and hence we will restrict B to integer values)
and futher investigate the noncommutative version of this
requirement.

In the language of group theory, considering $j$ integer 
means considering representations of the group $SO(2,1)$,
rather than just of its algebra, see ref.\cite{witten} 
for an illuminating discussion. 
Thus this requirement is
immediatly transferred to the noncommutative case, where
it implies requiring the quantum numbers labeling the
representations to be integers (or half-integers, if we
allow for $\pm$ signs). We will contrast the relevance of the
more general (non integer) representations of the algebra and the
richer set of spectrum  it gives with the restricted set applicable
for Riemann surfaces etc. in our analysis.  

This discrete part of the spectrum, which we will call Landau Levels,
comprises the eigenvalues
\beq
E_j=j(1-j),
\eeq
with $j=B-l$ up to the maximal $l=B-1$, each having a degeneracy
corresponding to the eigenvalues $J_3=(B-l)+n$, with $n$ any nonnegative
integer.

The corresponding wavefunctions are
\beqa
\Psi_l^{(n)} &=& (\pa -(B/2-1)\pa ln g_{z\bar z})\cdot
 (\pa -(B/2-2)\pa ln g_{z\bar z})\cdot \non \\
            &\cdots &(\pa -(B/2-l)\pa ln g_{z\bar z})\
            (i+z)^{2l}\Psi_0^{(n)}.
\eeqa
Besides the above discrete levels, there is a continuum
spectrum with nonnegative values for $E$.

The above results can be cast in an algebraic form, by
making use of the invariance group of $AdS_2$, that is $SO(2,1)$.
The $AdS_2$ manifold is conveniently described by embedding it
in flat Minkowski manifold with coordinates $x_1,x_2,x_3$
with the constraint:
\beq
x\circ x= x_1^2+x_2^2-x_3^2=-1, \label{xsq}
\eeq
where we have defined $V\circ W\equiv V_1W_1+V_2W_2-V_3W_3$ for
two vectors $V$ and $W$.

The $SO(2,1)$ generators $J_1,J_2,J_3$ satisfy the commutation
relations:
\beq
[J_1,J_2] = -iJ_3 \hskip.5truecm [J_2,J_3]=iJ_1 \hskip.5truecm
[J_3,J_1]=iJ_2, \label{Jcomm}
\eeq
\beq
[J_1,x_2] = -ix_3 \hskip.5truecm [J_2,x_3]=ix_1 \hskip.5truecm
[J_3,x_1]=ix_2.   \label{xcomm}
\eeq
We are considering here the standard commuting operators for $x$,
therefore
\beq
[x_l,x_n]=0.
\eeq
The relation with the previous formalism in terms of operators in complex
coordinates are:
\beq
x_1=i{z+\bar z \ov z-\bar z} \hskip.5truecm x_2=i{1-z\bar z \ov z-\bar z}
    \hskip.5truecm x_3=i{1+z\bar z \ov z-\bar z},
\eeq
and
\beqa
J_1 &=& i(z\pa + \bar z \bar \pa) \non \\
 J_{3,2} &=& -{i \ov 2}((1 \pm z^2)\pa +(1 \pm \bar z^2)\bar\pa
            \pm B(z-\bar z)). \label{Ji}
\eeqa

We can verify that
$
x\circ J=-B
$
and that
$
H= J\circ J
$,
therefore  eq.(\ref{bound}) tells us that   $J\circ J +B(B-1) \geq 0$.

It is well known \cite{book} that the unitary representations of
SO(2,1) algebra
are of two kinds:
the discrete ones $D^{\pm}_j$, in which $J\circ J =j(1-j)$
and $J_3 =\pm (j,j+1,..,j+n,...)$ with $j\geq 1/2$, (but restricted to
positive integer or half integer, if we look for representation of the
group instead)
and $n$ nonnegative integer,
and the continuum ones $C_j$ in which $J\circ J $ is real positive.

Therefore we find that the Landau Levels, we have obtained correspond to
$D^{+}_j$ with $j\leq B$.

\vskip.5cm

Note that if the surface is more in general described by the constraint 
$x\circ x= x_1^2+x_2^2-x_3^2=-r^2$ (previously $r=1$),
then the metric is $ds^2= r^2{ dx^2+dy^2 \over y^2}$ 
and the Hamiltonian (\ref{hamiltonian}) is now
$H= r^2 J\circ J$ while $x\circ J=-rB$. Thus the discrete
spectrum is $E_j=r^2 j(1-j)$, with the same $j=B-l$ as for
$r=1$.

It is convenient to redefine the Hamiltonian to be  
\beq
H_0=J\circ J,  \label{hamiltonian0}
\eeq  
for generic $r$, keeping the same definition $x\circ J=-rB$,
so that the discrete spectrum is always  
$E_l=-(B-l)(B-l-1)$. We will keep the Hamiltonian of 
eq.(\ref{hamiltonian0}) in the generalization to the 
noncommutative surface described in the next Section.

\section{Landau Levels in the noncommutative $AdS_2$}

\subsection{Definition of the problem}

In order to define the noncommutative $AdS_2$, we
introduce a
set of noncommuting coordinates $R_j$ with the $SO(2,1)$ algebra
as relevant noncommuting rules:
\beq
[R_1,R_2] = -iR_3 \hskip.5truecm [R_2,R_3]=iR_1 \hskip.5truecm
[R_3,R_1]=iR_2, \label{Rcomm}
\eeq
\beq
[J_1,R_2] = -iR_3 \hskip.5truecm [J_2,R_3]=iR_1 \hskip.5truecm
[J_3,R_1]=iR_2, \label{JRcomm}
\eeq
where the $J_i$ are the $SO(2,1)$ generators satisfying the algebra
eq.(\ref{Jcomm}).

Now, instead of requiring $x_1^2+x_2^2-x_3^2=-r^2$
which describes $AdS_2$ in the commutative case,
we require a fixed negative
value for the Casimir $ R\circ R\equiv R_1^2+R_2^2-R_3^2$.
We know from the $SO(2,1)$ representation theory \cite{book}
that such a negative Casimir is of the form
$ R\circ R = r(1-r)$. 
Of the two discrete representations ($D_r^{\pm}$ distinguished by positive
or negative $R_3$) we choose $D_r^{+}$, and then
$R_3=r,r+1,...$ and so on.


We still maintain the Hamiltonian to be:
\beq
H_0= J\circ J\equiv J_1^2+J_2^2-J_3^2,  \label{noncommham}
\eeq
as it is formally in the commutative case.

We note that the system is described by two mutually commuting
$SO(2,1)$ algebrae, $K_i=J_i-R_i$ and $R_i$.

The relevant decompositions are
(see ref.\cite{thesis} for a useful summary on combining $SO(2,1)$
representations):

\beq
D_k^{+}\otimes D_r^{+}=\sum_{m=0}D_j^{+}, \ \   j=k+r+m,\  m \  integer, \label{pp}
\eeq
 and
\beq
D_k^{-}\otimes D_r^{+}=(\sum_{j}^{|r-k|}D_j^{\pm} ,\ j=|r-k| ~ mod(1)+n,\  n\ integer) 
\oplus \int C_j.  \label{mp}
\eeq
with $D_j^+$ for $r> k$ and $D_j^-$ for $k> r$.

The other relevant formula is:
\beq
C_k\otimes D_r^{+}=(\sum_{j=r+n}D_j^{+},\ n=0,1,\dots )\oplus \int C_j.  \label{cp}
\eeq

In the commutative case we fixed $B=-(x\circ J)/r$ and studied
the spectrum with this additional constraint.
Now we must analogously decide what to fix to represent
the constant magnetic field.

{\bf 1)} We may follow the philosophy of Nair and Polychronakos ref.\cite{NP}
  and fix the value of the two Casimirs.
 Since $R\circ R$ is fixed by definition,
this amounts to parameterize the magnetic field by the choice of
$K\circ K$. 

Note that the approach of ref.\cite{NP} is such that 
it allows a redefinition of the Hamiltonian by an overall constant
which can be positive or negative depending on the range of the
parameters, in particular of the magnetic field. 

Our approach is to keep always the definition of the Hamiltonian
as in eq.(\ref{noncommham}).

{\bf 2)} Alternatively we we may stick to the choice similar to the
one in the commutative case, and use the commuting set of observables
$J\circ J,\ K\circ K,\ R\circ R,\ J_3$
to keep fixed
$$
K\circ K-J\circ J= R\circ R-2R\circ J.
$$
That is, like in the commutative case, we define and keep fixed 
$B\equiv -(x\circ J)/r$. 
This, we consider, is  more appropriate to our definition of the problem,
in keeping with the Hamiltonian of eq.(\ref{noncommham}).

In this case the limit $r \to\infty$ (at fixed $B$) is expected to reproduce the
physics of the Landau Levels on the commutative $AdS_2$ surface:
in fact by defining $x_i\equiv R_i/r$ we get approximate commuting
coordinates. 

Let us explore the resulting spectrum for both the choices.

Following the discussion of the previous Section, we may require the 
eigenvalue $j$ (and also $r$ for consistency) to be integer or half-integer.
This will be probably relevant for the setting of the Landau Levels problem
on a noncommutative Riemann surface, which is still to come.
Since the derivation is essentially the same, we consider 
$j$ to be real as the  general case. Requiring integer or half-integer $j$
would simply imply to discard the levels in which $j$ is not so;
the allowed levels would be much sparser then in the real $j$ case,
also depending on a fine tuning of the values of $r$ and $B$. 

\subsection{Case 1}

Let us start with  $K\circ K > 0$. Since $J_i=K_i+R_i$, the resulting spectrum of the
Hamiltonian is obtained from eq.(\ref{cp}). 
We get a continuum nonnegative part of the spectrum $C_j$
and an unbounded
discrete spectrum $D_j^{+}$ with $J\circ J =j(1-j)$, $j=r,r+1,...$
up to infinity;
therefore the Hamiltonian (\ref{noncommham}) is unbounded from below and above.

If $K\circ K < 0$, we have to consider two cases, corresponding
to the representations $D_k^{-}$ and $D_k^{+}$.

For $D_k^{-}$, the relevant decomposition is eq.(\ref{mp}).
We get a nonnegative continuum spectrum $C_j$ as well as a finite discrete
set of Landau Levels, $D_j^{\pm}$, with $J\circ J = -j(j-1)$; $j= |r-k|, |r-k|-1,...|r-k|mod(1)$ and hence bounded from below.

For $D_k^{+}$ the relevant decomposition is eq.(\ref{pp}),
giving thus again an unbounded negative discrete spectrum,
and therefore the Hamiltonian (\ref{noncommham}) is unbounded from below.

\subsection{Case 2}

Here we do not choose a particular value for
$K\circ K$ and therefore the spectrum is composed of various
parts, which must be consistent with the magnetic constraint
\beq
K\circ K-J\circ J=N\equiv -r(r-1)+2rB.  \label{magcon}
\eeq
We take $N$ to be positive or negative. There are still several
possibilities in the parameter space.

\vskip.5cm

\def\ha{{1\over 2}}

{\bf a)} Since $N=-r(r-1)+2rB$ and we would like to discuss in particular the case $r$ large with $B$ fixed
(because in this limit we recover the commutative $AdS_2$), we begin by assuming
$N<0$. Let us define $M=-N>0$: 
the magnetic constraint reads
\beq
J\circ J-K\circ K=M\equiv r(r-1)-2rB .  \label{mag}
\eeq
In particular, when both $J\circ J$ and $K\circ K$ are in the discrete part of the spectrum,
the constraint reads
\beq
(k-\ha )^2-(j-\ha )^2=M.  \label{magd}
\eeq 

We analyze different ranges for $M$. 
\vskip1cm

{\bf ai)} $\sqrt M \geq r$: this means $B<-{1\over 2}$. 
\vskip.5cm

We find that  $D^+_k$ cannot occur because (\ref{pp}) is incompatible with the 
constraint (\ref{mag})  which would require 
$(j-\ha )^2-(k-\ha )^2=-M$ which is impossible since $j>k$.
\vskip.5cm

The case $D^-_k$ is possible. 

First of all, from (\ref{mp}) we can have the continuum $C_j$,
with $j$ real positive and $j^2=M-k(k-1)$; since $k$ is an arbitrary positive parameter, we get the part of the continuum
spectrum for $j\leq M$. 

If $k>r$ we can also have a discrete part of the spectrum:
in this case from (\ref{mp}) we have $D^-_j$.  The constraint (\ref{magd}) is solved by:
\beqa
j_l&=&(|{M\over n_l}-n_l|+1)/2  \label{magj} \\
k_l&=&({M\over n_l}+n_l+1)/2   \label{magk}
\eeqa
We can always choose to restrict $n_l\leq \sqrt M$. Since from (\ref{mp}) we have
$j_l=k_l-r -l$ we get $n_l=r+l$, with $l$ a nonnegative integer; this is consistent
with $\sqrt M \geq r$. Since the minimum (maximum) possible $j_l$ is obtained for the 
maximum (minimum) possible $n_l$, we get the following discrete part of the spectrum
$J\circ J=-j_l(j_l-1)$:
\beq
\ha\leq j_l=({M\over r+l}-r-l+1)/2\leq ({M\over r}-r+1)/2.  \label{kgr}
\eeq
with $l$ integer restricted by $r+l\leq \sqrt M$.

If $r>k$ there is no discrete spectrum for $\sqrt M \geq r$. 

In fact we would have from (\ref{mp})
$j_l=r-k_l-l$; by using (\ref{magj}, \ref{magk}) this gives $n_l={M\over r-1-l}\leq M$.

Now $n_l(min)={M\over r-1}$ corresponding to $j_l(max)=(r-{M\over r-1})/2$,
whereas $j_l(min)\geq j_l(n_l=\sqrt M)=\ha$. 

Therefore, this is possible iff
$\ha\leq  (r-{M\over r-1})/2$ which implies $r\geq \sqrt M+1$.
\vskip.5cm

Finally, the case $C_k$ is also possible. It gives the part of the continuum spectrum
for $j\geq M$, since $j^2=M+k^2$ with $k$ arbitrary. We do not get a discrete part from it
 because, from (\ref{mag}), this would require $-j(j-1)=M+k^2$, which is not possible. 
\vskip.5cm

Summarizing: for $ \sqrt M>r$ we get the entire continuum spectrum ($0\leq j^2\leq \infty$) and the discrete part
of the spectrum reported in eq.(\ref{kgr}).
\vskip1cm 

{\bf aii)} $r-1\leq \sqrt M\leq r$: this means $-\ha\leq B\leq\ha -{1\over 2r}$.
\vskip.5cm

From the analysis of the case ai) we conclude that in this case we have only the continuum spectrum.
\vskip1cm

{\bf aiii)} $\sqrt M\leq r-1$: this means $ \ha -{1\over 2r}\leq B $ (and also $B<{r-1\over 2}$ in order
to have $M>0$).
\vskip.5cm

In this case from the analysis of the case ai) we conclude that we have the continuum spectrum,
and the following discrete part of the spectrum $J\circ J=-j_l(j_l-1)$, 
coming from the the representation $D^+_k$
and $r>k$:
\beq
\ha\leq j_l=(r-l-{M\over r-1-l})/2\leq (r-{M\over r-1})/2.  \label{rgk}
\eeq
with $l$ integer restricted by ${M\over r-1-l}\leq \sqrt M$.

\vskip1cm

Let us now consider the above results in the commutative limit $r\to\infty$ at fixed $B$,
looking at the lowest part of the spectrum. 

By taking the limit of eqs.(\ref{kgr}) and (\ref{rgk}),
and keeping the integer $l$ fixed 

\noindent
(note that ${M\over r-1-l}\sim r-2B+l$ and that ${M\over r+l}\sim r-1-2B-l$), 

\noindent
we find for $|B|>\ha$ the approximate discrete spectrum $J\circ J=-j_l(j_l-1)$ with:
\beq
j_l\sim |B|-l
\eeq
which is indeed the same result as for the commutative $AdS_2$.
In the Figs.1,2 we compare the discrete spectrum for the commutative and 
noncommutative case. 

For $|B|<\ha$ we only get
the continuum spectrum.

\begin{figure}[hbt]
\label{fig1} \vskip -0.5cm \hskip -1cm
\centerline{\epsfig{figure=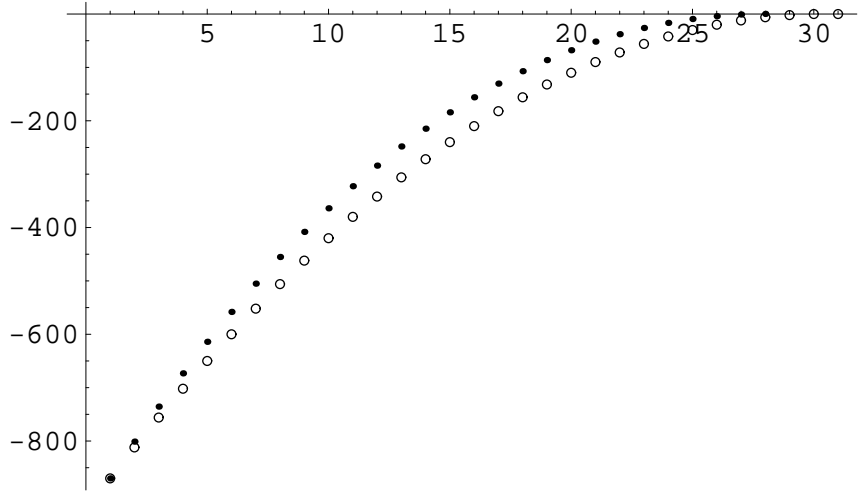,height=7truecm}}
\caption{\footnotesize Plot of the discrete levels $H_0=-j_l(j_l-1)$ as 
a  function of $l$ for the case 
$r=150$, $B=-30$ (full points), see eq.\ref{kgr}, compared with  the levels of the 
commutative case (open circles).}
\end{figure}

\begin{figure}[hbt]
\label{fig2} \vskip -0.5cm \hskip -1cm
\centerline{\epsfig{figure=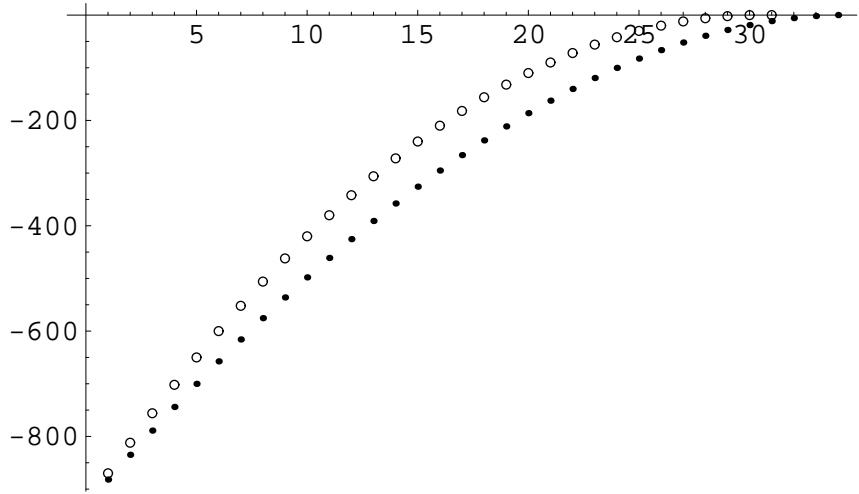,height=7truecm}}
\caption{\footnotesize Same as Fig.1 but  for the case
$r=150$, $B=30$, see eq.\ref{rgk}.}
\end{figure}

\vskip1cm

{\bf b)} Let us turn now for completeness to study the case 
$$
N\equiv -r(r-1)+2rB>0
$$
 which implies $B$ large in the commutative limit $r\to\infty$.

The magnetic constraint now reads:
\beq
K\circ K-J\circ J=N>0.     \label{magb}
\eeq
The analysis parallels the one done for the case a).
When both $J\circ J=-j(j-1)$ and $K\circ K=-k(k-1)$ are in the discrete spectrum this constraint
reads:
\beq
(j-\ha )^2-(k-\ha )^2=N,  \label{magbd}
\eeq
which can be solved by writing:
\beqa
j_l&=&({N\over n_l}+n_l+1)/2  \label{magjb} \\
k_l&=&({N\over n_l}-n_l+1)/2 .  \label{magkb}
\eeqa
With no loss of generality we have assumed $n_l\leq N$.

Also here we consider the three ranges of $N$.
\vskip1cm

{\bf bi)} $\sqrt N>r$. This means $B>r-1/2$.
\vskip.5cm

Now we can have $D^+_k$ since the decomposition (\ref{pp}) is allowed,
implying $n_l=r+l\leq \sqrt N$, with $l$ nonnegative integer. 

The maximum 
$j_l(max)=({N\over r}+r+1)/2$ is obtained for $n_l(min)=r$, whereas the minimum
$j_l(min)\geq \sqrt N+\ha$ corresponds to the maximum possible $n_l$. 

We thus get
a discrete part of the spectrum
$J\circ J=-j_l(j_l-1)$:
\beq
\sqrt N+\ha\leq j_l=({N\over r+l}+r+l+1)/2\leq ({N\over r}+r+1)/2.  \label{ppb}
\eeq
with $l$ nonnegative integer restricted by $r+l\leq \sqrt N$.
\vskip.5cm

Instead $D^-_K$ is not allowed: from the decomposition (\ref{mp}) we find that all the possibilities
are ruled out, since $D^-_j$ would imply $j\leq k-r$ contradicting the constraint (\ref{magb})
which gives $j>k$,  and the same constraint (\ref{magb}) would require for $C_j$ that
$j^2=-k(k-1)-N$ which is nonsense. 

As for the last possibility $D^+_j$, impying $r>k$,
the analysis is slightly longer: from the parametrization (\ref{magjb})  and (\ref{magkb}) we get
$n_l={N\over r-1-l}\leq \sqrt N$ (with nonnegative integer $l$) which in turn implies $\sqrt N\leq r-1$,
which is outside the range of $N$ considered here.
\vskip.5cm

We can also have $C_k$. The relevant decomposition is eq.(\ref{cp}) from which we get the
continuum spectrum $C_j$, that is $J\circ J=j^2$, with any value for $j$ since $k$ is arbitrary and $j^2=k^2-N$.

Moreover, from eq.(\ref{cp}) we also get another part of the discrete spectrum 
$D^+_j$, that is $J\circ J=-j_l(j_l-1)$, with
$j_l=r+l$, with $l$ nonnegative integer. 

The magnetic constraint (\ref{magb}) gives now:
$$
(j_l-\ha )^2=N+{1\over 4}-k^2 \Rightarrow l=\ha +\sqrt{N+{1\over 4}-k^2} -r.
$$
Since $k$ is arbitrary, this gives a possible range $0\leq l\leq \ha +\sqrt{N+{1\over 4}} -r$.
\vskip.5cm

Summarizing, we find for $\sqrt N>r$ the continuum spectrum  $J\circ J=j^2$ for any $j$,
and two parts of the discrete spectrum $J\circ J=-j_l(j_l-1)$, namely the part described in
eq.(\ref{ppb}) and another part in which 
\beq
r\leq j_l=r+l\leq \ha +\sqrt{N+{1\over 4}} . 
\eeq
Since $l$ is zero or integer, the two parts do not overlap.    
 \vskip1cm

{\bf bii)}  $r-1< \sqrt N\leq r$. This means $r-3/2+1/2r <B\leq r-1/2$.
\vskip.5cm

From the analysis of the case bi) we conclude that in this case we have only the continuum spectrum.
\vskip1cm

{\bf biii)}  $\sqrt N\leq r-1$. This means $r/2-1/2<B \leq r-3/2+1/2r$.
\vskip.5cm

From the analysis of the case bi) we see that $D^+_k$ is not allowed, whereas $D^-_k$ 
is allowed for $r>k$,  giving the discrete spectrum $D^+_j$, while the continuum $C_j$
is forbidden by the constraint (\ref{magb}). 

The eigenvalues for $j_l$ are of the form of 
eq.(\ref{magjb}) with $n_l={N\over r-1-l} $ with the nonnegative integer $l$ bounded by
$n_l\leq \sqrt N$. 

The maximum $j_l$ corresponds to the minimum $n_l$, that is for $l=0$,
and the minimum $j_l$ is obtained from the maximum $n_l\leq \sqrt N$ implying
$\sqrt N+\ha \leq j_l$. 

We thus get the discrete spectrum $J\circ J=-j_l(j_l-1)$ with
\beq
 \sqrt N+\ha\leq j_l=({N\over r-1-l}+r-l)/2\leq ({N\over r-1}+r)/2,  \label{disfin}
\eeq
with the nonnegative integer $l$ bounded by $l\leq r-1-\sqrt N$.
\vskip.5cm

Finally we can have $C_k$. From the analysis of the case bi) we see that we do not get 
here discrete spectrum, but only the continuum part $C_j$, with $j^2=k^2-N$ which is
always possible for any $j$.
\vskip.5cm

Summarizing, we find that for  $\sqrt N\leq r-1$ the continuum spectrum  $J\circ J=j^2$ for any $j$,
and the discrete spectrum  $J\circ J=-j_l(j_l-1)$ described in eq.(\ref{disfin}).

\vskip0.5cm
After this paper appeared on the net, we received a paper (hep-th/0201070)
by B.Morariu and A.P.Polychronakos, which implied a contrast with our
results. The present revised version
clarifies some points raised by hep-th/0201070.

\vskip1.0truecm

\noindent {\bf Acknowledgment}\\
RR would like to thank the High Energy Group of the Abdus Salam ICTP
for the support and the warm hospitality during the course of this work.

\end{document}